\documentclass[aps,twocolumn,superscriptaddress,showpacs,pre,floatfix,nofootinbib]{revtex4-1}
\usepackage{graphicx}
\usepackage{amssymb,mathtools,float,multirow}
\usepackage{amsmath,siunitx}
\usepackage{hyperref,multirow}
\hypersetup{colorlinks=true, citecolor=red, linkcolor=blue}
\usepackage{comment,color}
\newcommand{\corr}[1]{{\leavevmode\color{black}#1}}
\newcommand{\corrb}[1]{{\leavevmode\color{black}#1}}

\newcommand{\corrd}[1]{{\leavevmode\color{black}#1}}
\newcommand{\corca}[1]{{\leavevmode\color{black}#1}}
\newcommand{\corrand}[1]{{\leavevmode\color{black}#1}}

\begin{document}
\title{On the compatibility of the IceCube results with an universal neutrino spectrum}

\author{A. Palladino}
\email{andrea.palladino@gssi.infn.it}
\affiliation{Gran Sasso Science Institute, L'Aquila, Italy}
\author{C. Mascaretti}
\email{carlo.mascaretti@gssi.it}
\affiliation{Gran Sasso Science Institute, L'Aquila, Italy}
\author{F. Vissani}
\email{francesco.vissani@lngs.infn.it}
\affiliation{Gran Sasso Science Institute, L'Aquila, Italy}
\affiliation{Laboratori Nazionali del Gran Sasso, Assergi (AQ), Italy}

\begin{abstract}
%In this paper we present an update of the natural parametrization, introduced in \cite{natpar}, of the neutrino mixing (on cosmic distances) matrix $P_{\ell\ell'}$, taking into account the most recent oscillation parameters. We use such natural parametrization to perform a single flavor analysis, at the end of which we point out the different constraints that come from theory (neutrino oscillations), from the lack of data at high energy (resonant events and double pulses), and from the observations at low energy (HESE and MESE).
\corrd{There is mounting evidence that the IceCube \corrand{findings} \corca{cannot be} described simply invoking  
a single power-law spectrum for cosmic neutrinos.}
\corr{%The aim of this paper is to analyze the compatibility of an universal spectrum of cosmic neutrinos with the recent measurements provided by the IceCube collaboration.
\corrand{We discuss which are the minimal modifications of the spectrum that are required by the existing observations 
and we obtain a universal cosmic neutrino spectrum, i.e.\ valid for  all neutrino flavors.}
%Our results are based on a minimal set of hypotheses: 
\corrand{Our approach to such task can be outlined in three points:}
1) we \corrd{rely on} the throughgoing muon analysis above 200 TeV and \corrand{on} the 
high-energy starting events (HESE) analysis below this energy, requiring the continuity of the spectrum; 2) we assume that cosmic neutrinos are subject to three-flavor neutrino oscillations in vacuum; 3) we \corrand{make no assumption on the astrophysical mechanism of production, except for no $\nu_\tau$ $(\overline\nu_\tau)$ component at the source}. We test our model using the information provided by HESE shower-like events and by the lack of double pulses and resonant events. We find that a two power-law spectrum is compatible with all observations. 
The model agrees with the standard picture of \corca{pion decay as a source of neutrinos}, and indicates 
a slight preference for a $p\gamma$ mechanism of production. 
\corrand{We discuss the tension between the HESE and the \lq\lq throughgoing muons\rq\rq \ datasets around few tens TeV, focussing on the angular distributions of the spectra.
The expected number of smoking-gun signatures of $\nu_\tau$-induced events (referred to as \emph{double pulses}) is quantified: in the baseline model we predict 0.65 double pulse events in 5.7 years. Uncertainties in the predictions are quantified.}}
\end{abstract}
\maketitle

{\footnotesize
\tableofcontents}

\begin{figure*}[t]
\includegraphics[width=.32\textwidth,keepaspectratio]{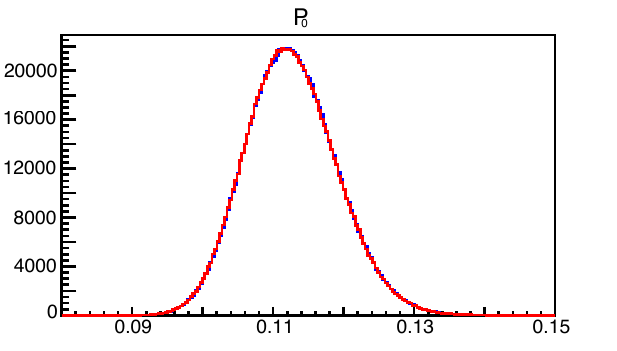}
\includegraphics[width=.32\textwidth,keepaspectratio]{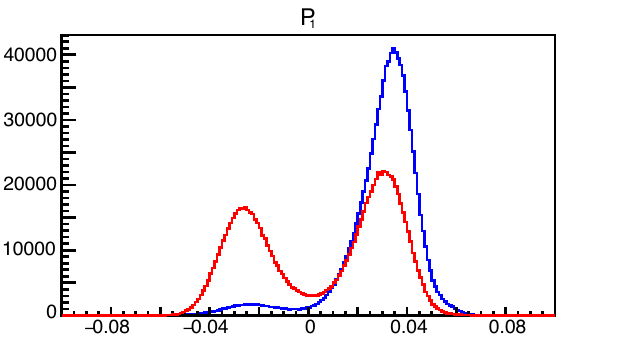}
\includegraphics[width=.32\textwidth,keepaspectratio]{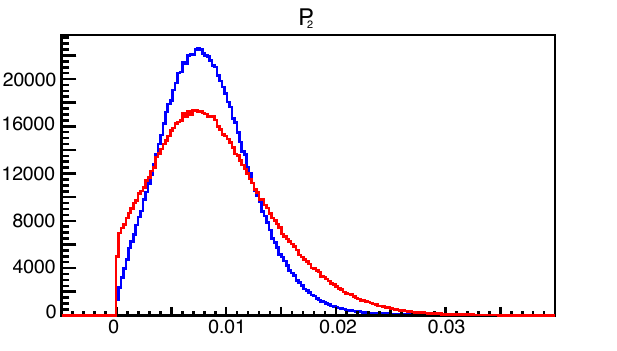}
\caption{\textit{Distribution of $P_0, \ P_1, \ P_2$ for normal \corrb{hierarchy} (blue) and for \corrb{inverted hierarchy} (red).}}
\label{fig:p012}
\end{figure*}

\section*{Introduction}
IceCube observed a new component of the neutrino spectrum, that exceeds the atmospheric neutrino flux above few hundreds TeV. \corrand{This new component extends, at least, up to few PeV and it has} an intensity close to the Waxman-Bahcall \corrand{upper} bound \cite{Waxman:1998yy}. 

This is one of the most exciting recent results in neutrino physics, even though we do not know which are the  sources of these neutrinos. The energy spectrum displays non trivial and even unexpected features, such that the aim of the present work is to investigate what is the consequence of a global interpretation of the IceCube findings. 

We base our analysis on a minimal set of hypotheses, namely:
\begin{enumerate}
\item \corrand{the spectrum is continuous and regular}, which is not only a simple mathematical requirement but also a reasonable assumption, as the existence of major discontinuities would require some specific motivation, that we do not have currently;
\item the  
cosmic neutrinos are subject to three-flavor neutrino oscillations, as recently prove\corr{d} by terrestrial experiments and observations;
\item the new population of cosmic neutrinos derives from \corrand{some unspecified} astrophysical mechanisms of production, \corrand{where} $\nu_e$ ($\bar{\nu}_e$) and $\nu_\mu$ ($\bar{\nu}_\mu$) are created at the source.
%, unless supposed otherwise, from the decay of unstable mesons (muons, charged pions and kaons) presumably formed by cosmic ray collisions close or at their sources. 
\end{enumerate}
Moreover, we \corrd{consider the most recent datasets obtained by IceCube, discussing the relevant backgrounds.} %assume the absence of major systematic errors in the current IceCube data analyses.

These hypotheses restrict significantly the overall shape of the spectrum. The hard power-law spectrum, that describes the induced muons up to a few PeV, can be extended \corca{to} low energy either by assuming a piecewise functional form or by adding a softer power-law component, but there is no tangible difference, as the resulting flux is quite constrained. This has direct implications \corrand{for} the physics of muon neutrinos events -- HESE tracks or throughgoing events. 
Neutrino oscillations \corrand{allow us to derive} the electron and tau neutrino spectra. 
Possible deviation from the standard pion decay scenario are analyzed. We show that there is a hint of a slight excess of electron neutrinos and antineutrinos, but this is not significant.  
Several tests of the ensuing physical picture are discussed, including 
%the low  energy track events, 
tau neutrino events (that are detectable), Glashow resonance events, examining their relation to the specific dataset or range of the neutrino spectrum.
\corrand{We examine the dependence of the predictions upon the specific dataset and upon the energy range of the universal spectrum.}

\section{Neutrino oscillations}
\corrand{In this section we update} the description of cosmic neutrino oscillations proposed in \cite{natpar}. This consists in the use of
%\corr{In this section} we present an update of the work originally done in \cite{natpar}, which consisted in the use of 
three \lq\lq natural\rq\rq \ parameters to describe the \corr{probabilities of oscillations of cosmic neutrinos}. After a brief review of \cite{natpar}, we discuss our updating procedure, which is based on the latest \corrd{results on} the oscillation parameters \cite{bari}.

\subsection{The parameters $P_0$,  $P_1$, $P_2$}
\corrd{The average \corca{survival}/oscillation probabilities in vacuum are given by:}
%\corr{The matrix elements are defined as follows:}
\begin{equation} P_{\ell\ell'}=\sum_{i=1}^n |U_{\ell i}^2| |U_{\ell' i}^2| \qquad \ell,\ell' = e,\mu,\tau \end{equation}
\corrd{where $\ell(\ell')$ denotes the neutrino flavor and $U$ is the standard mixing matrix.}

The approach of Palladino \corr{and Vissani in }\cite{natpar} to compute the average survival/oscillation probabilities, \corrd{of cosmic neutrinos in vacuum,} is based on two \corr{simple considerations}:
\begin{itemize}
\item the matrix $P$, \corr{containing the probabilites of oscillation $P_{\ell \ell'}$} is symmetric under \corrd{the exchange of the flavor indices} $\ell \leftrightarrow \ell'$ 
\item \corr{the elements of the mixing matrix must obey the condition}
$\sum_\ell P_{\ell \ell'}=1$
\end{itemize}
For these reasons, the number of independent parameter is $n(n-1)/2$, where $n$ is the number of neutrinos. For $n=3$ we have just 3 independent parameters. Calling them $P_0,P_1$ and $P_2$, we can
\corr{write the probabilities as the following matrix}:
\corr{\begin{equation}
P=\frac{\mathbb{I}}{3}+
\begin{pmatrix}
2P_0  &-P_0+P_1 & -P_0-P_1 \\
&  P_0/2-P_1+P_2 &  P_0/2-P_2 \\
 & & P_0/2+P_1+P_2 \\
\end{pmatrix}
\end{equation}
The expressions of these  
parameters in terms of the conventional oscillation parameters 
(3 mixing angles and one CP violating phase) are:
%and thus they worked out a \lq\lq natural\rq\rq parametrization with just three independent parameters, namely:
\begin{align}
P_0 &=\frac{1}{2} \left[(1-\epsilon)^2 \left( 1-\frac{\sin^2 (2\theta_{12})}{2}\right)+\epsilon^2 -\frac{1}{3} \right] \label{eq:p0}\\[2mm]
P_1 &=\frac{1-\epsilon}{2}\left(\gamma\cos2\theta_{12}+\beta\frac{1-3\epsilon}{2} \right)\label{eq:p1}\\[2mm]
P_2 &=\frac{1}{2}\left[\gamma^2 + \frac{3}{4}\beta^2 (1-\epsilon)^2 \right] \label{eq:p2}
\end{align}
where
\begin{align*}
\epsilon &= \sin^2\theta_{13} \qquad \alpha = \sin\theta_{13}\cos\delta\sin2\theta_{12}\sin2\theta_{23} \\[1mm] \beta &= \cos2\theta_{23} \qquad \gamma = \alpha - \frac{\beta}{2}\cos2\theta_{12}(1+\epsilon)\end{align*}
%\corr{With this choice of the parameters we have that $P_0 \gg P_1, P_2$.}
}

\subsection{The oscillation probabilities of cosmic neutrinos}
\label{sec:nuosc}
The values of the conventional oscillation parameters are given in \cite{bari}; \corr{in the following table we report their best fit values and the 68\% confidence level interval, \corrb{denoting with NH the normal hierarchy/ordering and with IH the inverted hierarchy/ordering.}}

\begin{table}[h!]
\corrb{\begin{tabular}{ccccc}
\hline
&&& \\[-2.5mm]
ordering & $\sin^2_{\theta_{12}}$ &$\sin^2_{\theta_{23}}$& $\sin^2_{\theta_{13}}$ & $\delta/\pi$ \\[.5mm]
\hline 
&&&&\\[-2.5mm]
\corrb{NH} & \multirow{2}{*}{$0.297^{+0.017}_{-0.016}$}  & $0.425^{+0.021}_{-0.015}$ &$0.022^{+0.001}_{-0.001}$ & $1.38^{+0.23}_{-0.20}$ \\[1mm]
\corrb{IH} & & $0.589^{+0.026}_{-0.022}$ &$0.022^{+0.001}_{-0.001}$ & $1.31^{+0.31}_{-0.19}$   \\[1mm]
\hline
\end{tabular}}
\end{table} 
\normalsize

%\begin{center} 
%\includegraphics[width=.45\textwidth,keepaspectratio]{opars_bari.pdf}
%\end{center}

The distributions of the oscillation parameters are sampled according to likelihood functions \corr{reported in figure 1} of \cite{bari}. \corr{This approach is necessary because the parameters $\sin^2\theta_{23}$ and  $\delta/\pi$ are not Gaussian distributed.} %for us this meant that, in the Monte Carlo with which we obtained the $P_i$ parameters, we 
%We used the relevant likelihood functions, as reported in \cite{bari}, to sample $\sin^2\theta_{23}$ and $\delta/\pi$, 
\corr{On the contrary, for $\sin^2\theta_{12}$ and $\sin^2\theta_{13}$ it is sufficient to use} Gaussian distributions, with mean the central value and with standard deviation the average of the errors quoted in the table above. 
Performing Monte Carlo extractions \corca{according to such procedure}, we obtain the distributions for $P_0$, $P_1$ and $P_2$ shown in figure \ref{fig:p012}; their \corr{best fit} values and 68\% CL interval\corr{s} are reported in table~\ref{tab:nat_pars}.
\begin{table}[t]
\caption{\textit{The best fit values and 68\% intervals that we obtained for the natural parameters.}} \vskip3mm
\label{tab:nat_pars}
\begin{tabular}{ccccc}
\hline
&&&\\[-2.5mm]
ordering & $P_0$ && $P_1$ & $P_2$  \\[.5mm]
\hline 
&&&&\\[-2.5mm]
\corrb{NH} & \multirow{2}{*}{$0.113\pm0.006$}  && $0.035^{+0.010}_{-0.012}$ & $0.008^{+0.005}_{-0.004}$ \\[1mm]
\corrb{IH} & && $0.029^{+0.010}_{-0.057}$ & $0.008^{+0.005}_{-0.006}$  \\[1mm]
\hline
\end{tabular}
\end{table}
%\corr{\huge{P2 in NO troppe cifre !! Meglio scriverla come tabella anziché immagine, così possiamo modificarla}}

%\normalsize

\corca{From the table it is clear} that:
\begin{equation}P_0 > P_1 > P_2\end{equation}
$P_0$ is the \corrd{largest} parameter, and also the one with the smallest uncertainty.
\corca{From the plot we see that} the parameter $P_2$ satisfies the condition $P_2>0$
consistently with equation \eqref{eq:p2}.
The asymmetric errors quoted in the table are such that 
the integral of the normalized \corca{distribution $\mathcal L_P$ of a generic parameter $P$} obeys the condition\corca{s}:
\corrand{\begin{equation}
\left\{ \begin{array}{l}
\displaystyle \int_{P_\text{BF}-\Delta P_-}^{P_\text{BF}+\Delta P_+} \mathcal{L}_P (t) \ dt = 0.68 \\[4ex]
\mathcal{L}_P (P_\text{BF}-\Delta P_{-}) = \mathcal{L}_P (P_\text{BF}+\Delta P_{+}) 
\end{array} 
\right.
\end{equation}}
where $P_\text{BF}$ is the best fit value and $\Delta P_+$, $\Delta P_-$ are the asymmetric errors.

\section{The IceCube dataset}
In this section we present two recent dataset\corca{s} provided by the IceCube collaboration \corr{after 6 years of data taking}: the throughgoing muon dataset and the high energy starting events (HESE) dataset. 

\corrd{\textbf{Notation:}} from here on we denote \corca{by} $\phi_\ell$ the flux of $\nu_\ell$ and of $\bar{\nu}_\ell$. Whenever we are \corrand{only} interested to the flux of neutrinos (or antineutrinos), we denote it \corca{by} $\phi_{\nu_\ell}$ (or $\phi_{\bar{\nu}_\ell}$). When the subscript is not present \corrd{$(\phi)$}, the all-flavor flux is considered.
\subsection{Throughgoing muons}
\label{sec:muon}
The IceCube collaboration \corca{acquired} data from 2009 to 2015, \corca{collecting} a sample of charged current \corca{events due to upgoing muon neutrinos; \corrand{due to the position of IceCube, the field of view, for this class of events,} is restricted to the Northern hemisphere} \cite{muon}.
The highest energy sample (with reconstructed energy above $\sim$ 200 TeV) corresponds to 29 events of this type; %, that are reported in Tab.\ref{tab:muoni6}. 
a purely atmospheric origin of them is excluded at more than 5$\sigma$ of significance. 
\corrd{The most energetic event corresponds to a reconstructed muon energy equal to 4.5~PeV.}

The corresponding cosmic muon flavor (neutrino and antineutrino) flux was obtained with a power-law fit to the data:
\begin{equation}
\frac{d\phi_\mu^{\mbox{\tiny data}}}{dE}=F_\mu \times \frac{10^{-18} }{\rm GeV \ cm^{2} \ s \ sr } \left(\frac{E}{\rm 100 \ TeV} \right)^{-\alpha}
\label{eq:muon}
\end{equation}
The parameters are $F_\mu=0.90^{+0.30}_{-0.27}$ and $\alpha=2.13 \pm 0.13$. This analysis is sensitive only to muon neutrinos and antineutrinos.
No correlation with known $\gamma$-ray sources was found by analyzing the arrival directions of these 29 events \cite{muon,bllac}.

\subsection{High Energy Starting Events}
\label{sec:hese}
%%%%%%%%%%%%%%%%%%%%%%%%%%%%%%%%%%%%%%%%%
%An important dataset collected by IceCube concerns the contained events, better known as high energy starting events (HESE). 
\corrd{The most recent data concern} 2078 days (5.7 years) of \corca{detection}. \corrd{This dataset includes} 82 HESE \cite{iceproceeding}: they have been classified in 22 tracks and 58 showers (2 of them are not classified being coincident events). %Among them, 16 events come from the Northern hemisphere, 37 events come from the Southern hemisphere and 1 event was detected with declination equal to zero. 
These events are characterized by a deposited energy larger than 30 TeV, and the most energetic HESE deposited an energy of 2~PeV into the detector. 
%Also other 2 events with deposited energy above 1 PeV have been detected.  

The flux attributed to astrophysical neutrinos is described, in first approximation, by an isotropic distribution and a power-law spectrum. The all-flavor flux is:
\begin{equation}
\frac{d\phi^{\mbox{\tiny data}}}{dE}= F \times \frac{10^{-18}}{\rm GeV \ cm^{2} \ s \ sr } \left(\frac{E}{\rm 100 \ TeV} \right)^{-\alpha}
\label{eq:hese}
\end{equation}
with $F=2.5 \pm 0.8$ and $\alpha=2.92^{+0.33}_{-0.29}$ \cite{iceproceeding}. We denote by $F$ the normalization of the all-flavor flux.

Although the bulk of HESE coming from the Southern sky suggest\corrd{s} a power-law spectrum with spectral index $\alpha\approx 2.9$, the subset of highest energy (above 200 TeV) HESE is in agreement with a much harder spectrum and, more precisely, follows the same distribution suggested by the throughgoing muons: 
see \corr{figure 6} of \cite{nugal1}
and \corr{figure 5} of \cite{muon}, and discussions therein.
In other words,  
%\begin{quote}
the flux of the highest energy HESE observed from the Southern sky is compatible with the same hard spectrum, $\alpha\approx 2$, suggested by throughgoing muons.

\section{Atmospheric background of HESE}
\label{sec:backhese}
Before continuing the discussion, it is important to recall what are the backgrounds for high energy neutrinos. \corrand{A precise knowledge of the different background sources is relevant for the correct identification of the astrophysical signal, that we perform in section \ref{sec:nuspec}.}

When cosmic rays collide with the terrestrial atmosphere, lots of mesons are produced: from pion decay (and from kaon decay, in smaller amounts) muons and neutrinos are produced, constituting the main source of background for high energy neutrino detection. We call these two sources of background as \corca{atmospheric} muons and conventional \corrd{neutrino} background.

Another contribution to the background is given by the decay of heavy, charmed mesons: the neutrinos which come from these decays are called ``prompt neutrinos''.

\subsection{Atmospheric muons}
Atmospheric muons, mainly generated by \corrand{pion decay}, have an energy spectrum $\propto E^{-3.7}$. This is due to the fact that, with \corca{increasing} energy, the probability that pions interact before decaying grows linearly \corrd{with} $E$. Since muons come from pion decay, their spectrum is steeper than the $E^{-2.7}$ spectrum of primary cosmic rays. This is an unavoidable source of background for the \corr{HESE analysis; \corrand{on the other hand}, it does not affect the througohgoing muon analysis, since atmospheric muons are absorbed crossing Earth.} It has been estimated by the IceCube collaboration that \corr{the} \corrd{the number of atmospheric muons, contributing to HESE background} after 5.7 years of exposure, is:
\begin{equation}
b_\mu= 25.2 \pm 7.3 
\end{equation}
According to \corr{table 4} of \cite{ice3yr}, 90\% of them ($23.0 \pm 7.3$) are identified as track-like events and 10\% ($2.2 \pm 0.7$) as shower-like events. \corr{This is due to the fact that a certain misidentification of tracks is possible from an experimental point of view.}

\subsection{Prompt neutrinos}
\label{sec:prompt}
Prompt neutrinos are produced in the decay of heavy mesons, which contain the charm quark (charmed mesons). \corr{These particles are highly unstable and decay before interacting, following the same $E^{-2.7}$ spectrum of primary cosmic rays.} 

To date, \corca{the contribution of prompt neutrinos to the IceCube dataset 
has not been yet identified}, although it is expected to exist: \corrd{see e.g.~\cite{ers,prompt2,prompt3}}. An upper limit has been set by the IceCube collaboration \cite{ice3yr}, while Palladino et al. \cite{nugal2} have calculated that their contribution to HESE is smaller than 3.5 events, in 4 years of exposure, at 90\% confidence level (CL). Scaling such estimate with the present exposure, we obtain that the contribution of prompt neutrinos is expected to be smaller than 5 HESE, at 90\% CL. 

Since at the time of the writing the best fit value of prompt neutrino events is 0, the probability density function (PDF) of prompt neutrinos \corr{can be reasonably} approximated by an exponential \corr{function}:
\begin{equation}
\mathcal{L}_{p}(b_p)=  \frac{1}{b_p^0} \exp \left(-\frac{b_p}{b_p^0} \right)
\end{equation}
with $b_p^0=2.17$.

According to \corr{table 4} of \cite{ice3yr} \corr{about} 20\% of prompt neutrinos produce track-like events, whereas \corr{about} 80\% of them produce shower-like events. 

\subsection{Conventional background}
Neutrinos produced in the decay of pions (and kaons, in smaller amounts) constitute the so called conventional background. These neutrinos \corca{have} an $E^{-3.7}$ \corca{energy} spectrum, for the same reason discussed in the case of atmospheric muons. 

The IceCube collaboration \cite{iceproceeding} has estimated that the contribution of atmospheric neutrinos (conventional plus prompt) to the HESE background is equal to:
\begin{equation}
b_{\pi k}+b_p= 15.6^{+11.4}_{-3.9}
\end{equation}

In order to isolate the contribution of conventional neutrinos, we have built the likelihood function $\mathcal{L}_{\pi k +p}(b)$ that \corca{reproduces} the \corr{best fit value} and the 68\% CL \corca{asymmetric} interval. % \corrd{on both sides, keeping into account the different errors}. 
We obtain the PDF of conventional neutrinos marginalizing over $b_p$:
\begin{equation}
\mathcal{L}_{\pi k}(b_{\pi k})= \int_0^\infty \mathcal{L}_{\pi k +p}(b_{\pi k}+b_p) \times  \mathcal{L}_{p}(b_p) \ d b_p
\end{equation}
\corr{Following this procedure,} the expected background from conventional neutrinos is equal to:
\begin{equation}
b_{\pi k}=14.7^{+10.8}_{-5.1}
\end{equation}
where we quote the best fit value and the 68\% CL interval obtained \corr{as described in sction \ref{sec:nuosc}}, i.e.\ using the condition that the integral of the normalized likelihood function is equal to 0.68 between $b_m$ and $b_M$ and $\mathcal{L}(b_m)=\mathcal{L}(b_M)$. This is a general procedure that we use for every asymmetric function \corr{from here on.} \corrd{We have verified that the same result is obtained performing a Monte Carlo extraction for the total background and for prompt neutrinos.}

According to \corr{table 4} of \cite{ice3yr}, 70\% of them ($10.3^{+9.1}_{-4.7}$) contribute to track-like events, whereas 30\% of them ($4.4^{+4.2}_{-2.0}$) contribute to shower-like events. The uncertainties on the expected number of showers and tracks %are such that they 
reproduce the total uncertainty when summed in quadrature. 

\subsection{Summary of backgrounds}
\label{sec:background}
We summarize the backgrounds relevant to the HESE analysis in table \ref{tab:backhese}.

 The expected number of \corrd{background} tracks in the HESE \corrd{dataset} is equal to $34.3^{+12.3}_{-8.7}$, as reported in \corr{table \ref{tab:backhese}}. This number is large\corca{r} \corrd{than the observed} 22 tracks. Moreover, we \corrd{expect} that also $\sim 20\%$ of cosmic neutrinos produce tracks \corrd{in the HESE dataset}, according to table 4 of \cite{ice3yr}. On the other hand, as discussed in \cite{hese}, the misidentification of some tracks, that could be identified as showers, could play an important role \corrd{for} this kind of analysis. In conclusion, since the track-like subset  is 
 supposedly dominated by the atmospheric background rather than by the signal, \textit{it is quite hard to extract useful information on $\phi_\mu$ from \corrd{HESE, and this is the reason why we do note use this} subset of data in our analysis.}

On the contrary, we include the tracks contained into the throughgoing muons dataset, since they are affected by the atmospheric background at the level of 30\%, as estimated in \cite{bllac}. Moreover, \corr{we repeat that} this kind of analysis is free from atmospheric muons, since they are absorbed into Earth. 

As a final remark, let us consider that the atmospheric background affects shower-like events, in the HESE dataset, at the level of 15\%. Indeed the expected number of showers, due to atmospheric background, is 
\begin{equation}
b_s= 8.8^{+4.0}_{-3.0}
\label{eq:back}
\end{equation}
We denote by $\mathcal{L}_s(b_s)$ the distribution function of this background. This number has been obtained using a Monte Carlo simulation and combining the showers expected from \corr{atmospheric} muons, conventional neutrinos and prompt neutrinos.

It is reasonable, therefore, to consider throughgoing muons and shower-like HESE in our analysis, due to their small atmospheric background. On the other hand, it is \corca{cautious} to neglect track-like HESE in the rest of the analysis, due to the huge atmospheric background, \corr{that does not allow to extract useful information on the astrophysical signal}.

\begin{table}[t]
\caption{\textit{Summary of the backgrounds expected in HESE analysis after 5.7 years of exposure.}} \vskip3mm
\label{tab:backhese}
\begin{tabular}{ccccc}
\hline
&&&&\\[-2.5mm]
& $b_\mu$ & $b_{\pi k}$ & $b_p$ & Sum \\[.5mm]
\hline 
&&&&\\[-2.5mm]
Tot. events & $25.2 \pm 7.3$ & $14.7^{+10.8}_{-5.1}$ & $<5.0$ at 90\% CL & \corrd{$43.1^{+12.9}_{-9.2}$} \\[1mm]
Tracks  & $23.0 \pm 7.3$ & $10.3^{+9.9}_{-4.7}$ & $<1.0$ & $34.3^{+12.3}_{-8.7}$ \\[1mm]
Showers  & $2.2 \pm 0.7 $ & $4.4^{+4.2}_{-2.0} $ & $<4.0$ & $8.8^{+4.0}_{-3.0}$ \\[1mm]
\hline
\end{tabular}
\end{table}

\begin{figure*}[t]
\includegraphics[scale=0.7]{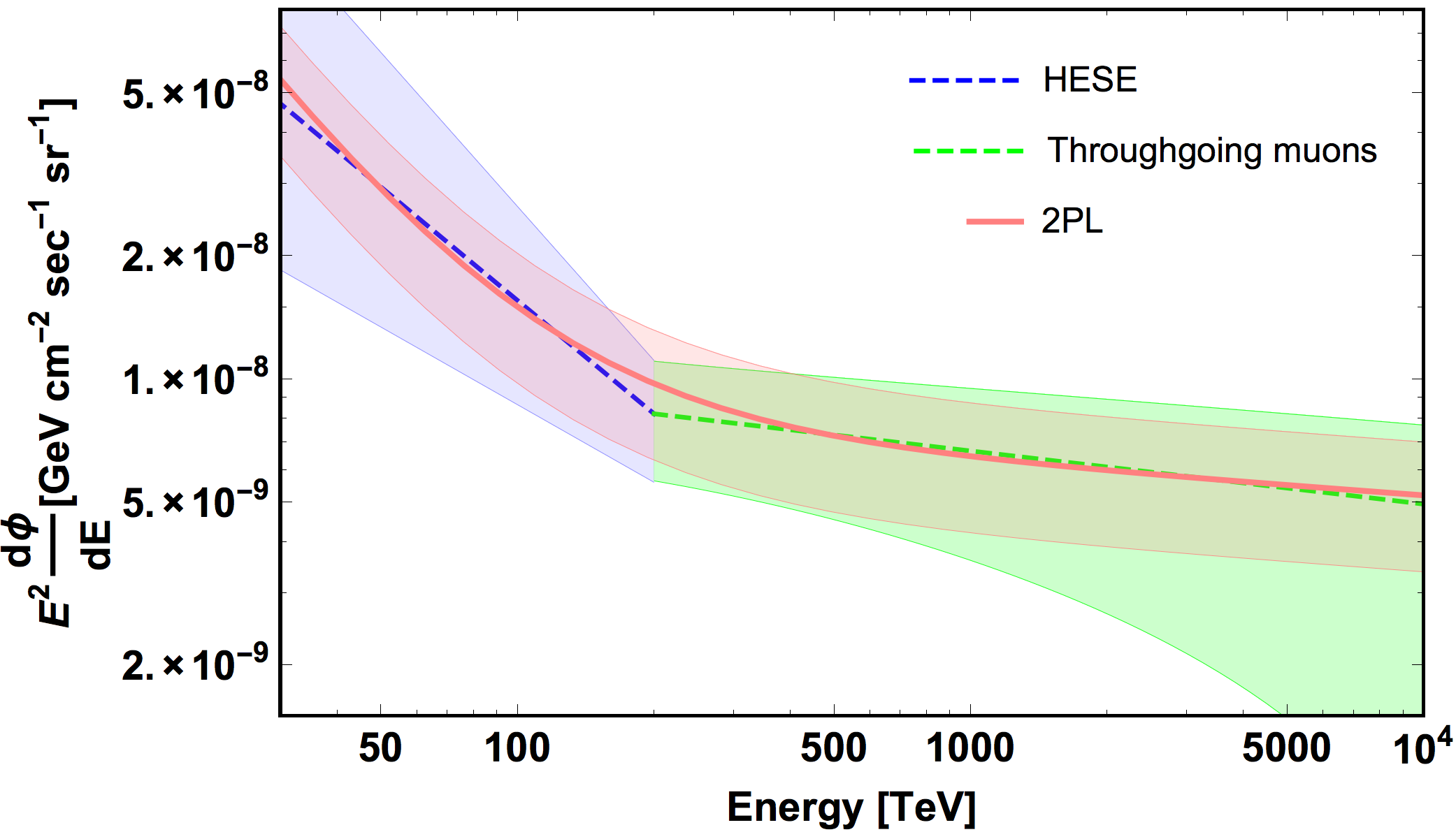}  
\caption{\textit{%On the left panel: the spectral indices of HESE (below 200 TeV) and throughgoing muons (above 200 TeV) with their uncertainty. The pink line represents the spectral index of the 2power law spectrum, as a function of the energy. On the right panel: 
The two power-law benchmark spectrum (2PL) as defined by Eqs.~\eqref{stdflux}, \eqref{caretta1}, \eqref{eq:likmu}, 
compared with the throughgoing muons flux (green band) 
and a flux with the slope suggested by HESE below 200 TeV (blue band)--compare with equation \eqref{brokenflux}.}}
\label{fig:brokenflux}
\end{figure*}

\section{The neutrino spectrum}
\corrand{In subsection \ref{muspec} we define the \lq\lq universal\rq\rq \ spectrum, starting from the muon neutrino spectrum. This kind of spectrum reconciles all the recent IceCube measurements. In subsection \ref{tau:theory} we evaluate the spectrum of tau neutrinos, showing that neutrino oscillations are sufficient to strongly constrain it. In subsection \ref{sec:nueantinue} we evaluate the spectrum of electron neutrinos and electron antineutrinos. In this case we analyze $\nu_e$ and $\bar{\nu}_e$ separately, since they produce different signals in the detector and, as a consequence, they are distinguishable.}
\label{sec:nuspec}
\subsection{The shape of muon neutrino spectrum}
\label{muspec}
Combining all the information provided by the IceCube collaboration with their different analyses, it is evident that the assumption of a single power-law model is not \corr{the best choice} to explain the present data. In several papers \cite{nugal1,nugal2,model3,model4,model5,model6,model7} this aspect has been emphasized, invoking the presence of at least two populations of high energy neutrinos with different energy spectra. 

In this paper %we do not investigate the origin of high energy neutrinos, but 
we test the compatibility of a two power-law spectrum with the observations \corr{(HESE showers, double pulses, resonant events)} and with the standard production mechanisms of high energy neutrinos, \corr{expected to occur} in astrophysical environments. %Therefore we assume that the minimal modification that can reconcile all the observations is a two power law spectrum. 

Above 200 TeV we \corrd{can rely on} the throughgoing muon analysis (green band in figure \ref{fig:brokenflux}), while below 200 TeV we \corrd{rely on} the HESE analysis (blue band in figure \ref{fig:brokenflux}), \corr{for the reasons discussed in the previous section.} In order to proceed, \corrd{we define the broken power-law flux in the following manner:
\corrand{\begin{equation}
\frac{d\phi_{br}}{dE}=\frac{N_\mu^{br} \ 10^{-18}}{\rm GeV \ cm^2 \ s \ sr} \left\{ \begin{array}{cc} 
E_{200}^{-2.13} \ \mbox{for} \ E\geq200 \mbox{ TeV} \\ [2.0mm]
E_{200}^{-2.92}  \ \mbox{for} \ E<200 \mbox{ TeV} \\
\end{array}  \right.
\label{brokenflux}
\end{equation}
%where $E_{200} = E/\SI{200}{TeV}$.
%The broken flux follows an $E^{-2.13}$ spectrum above 200 TeV and an $E^{-2.92}$ below 200 TeV, 
where $E_{200}=E/\rm 200 \ TeV$} and the normalization at 200 TeV is $N_\mu^{br}=0.206$ (in units of equation \eqref{brokenflux}); this value corresponds to the normalization of the throughgoing muons flux at 200 TeV, using the best fit values. The choice of the break at 200 TeV represents:
\begin{itemize}
\item the minimal modification that reconciles the throughgoing muon and the HESE dataset;
\item the most conservative choice, since the energy threshold of the throughgoing muon analysis is about 200 TeV.
\end{itemize}}

\corrd{Now, we define} our \lq\lq benchmark\rq\rq \ two power-law flux $\phi_\mu$ for the muon neutrino plus antineutrino spectrum as follows:
\begin{equation}
\frac{d\phi_\mu}{dE}=\frac{N_\mu}{2} \frac{10^{-18}}{\rm GeV \ cm^2 \ s \ sr} \left( E_{100}^{-\alpha}+E_{100}^{-\beta} \right)
\label{stdflux}
\end{equation}
\normalsize
where $E_{100} = E/\SI{100}{TeV}$.
Thanks to the prefactor $N_\mu /2$, the normalization $N_\mu$ denotes directly the normalization of the two power-law flux at 100 TeV. \corrd{The choice of the normalization at 100 TeV reproduces, reasonably well, the behavior of the broken power-law flux. The value can be slightly different but we have verified that choosing 150 TeV or 200 TeV the analyses proposed in the next sections are not affected \corca{appreciably}.}

In order to determine the parameters $N_\mu$, $\alpha$, $\beta$ of equation \eqref{stdflux}, we define a \corr{\lq\lq distance\rq\rq}  between this benchmark flux and the broken power-law flux $\phi_{br}$, i.e.\ the flux suggested by the data. The distance between the two functions is defined as follows:
\begin{equation}
d(N_\mu,\alpha,\beta)= \int_{\SI{30}{TeV}}^{\SI{10}{PeV}} \frac{|\phi_\mu(E,N_\mu,\alpha,\beta) -\phi_{br}(E) |}{\phi_{br}(E)}  \ d \log E
\end{equation}
Such distance is minimized by the following set of values:
\begin{equation}
N_\mu= 1.5 \ \ \ \alpha=2.08 \ \ \ \beta=3.5
\label{caretta1}
\end{equation}
Since the normalization of the throughgoing muon flux is known with an uncertainty of about $30\%$, we take it into account considering that
\begin{equation}
N_\mu= 1.5 \pm 0.5
\label{eq:likmu}
\end{equation}
The two descriptions of the fluxes are presented in figure \ref{fig:brokenflux}. 

%\huge
%\corrd{CONTINUA DA QUI}
%\normalsize

Let us recall that \corrd{assuming three flavor neutrino oscillations and the same mechanism of production for all cosmic neutrinos,} %from standard production mechanisms and from neutrino oscillations 
we expect that the shape of neutrino spectra is the same for all flavors, and only their normalization is expected to be different.
For this reason, we refer to assumption with the terminology: universal spectrum of neutrinos.

In figure \ref{fig:brokenflux} we see that the sum of the two power-law fluxes (pink band), with spectral indexes $\alpha=2.08$ and $\beta=3.50$, reproduces well the $\sim E^{-2.92}$ behavior at low energy and the $\sim E^{-2.13}$ behavior at high energy, \corr{within the uncertainties on the spectral index and on the normalization}. %Therefore, we assume that the flux of cosmic neutrinos is in agreement with the shape suggested by HESE below 200 TeV, i.e. in the energy region not measured using throughgoing muons. 

It is important to remark that we assume 
the shape of the spectrum suggested by the low energy HESE data, but we do not yet use 
the normalization suggested by the same data. In fact, HESE data refer to an all-flavor analysis, but  
the flavor partition of the neutrinos is dictated by the mechanism of production, that to date is unknown. Therefore, 
we include the information on HESE in the analysis by adopting the following procedure:
\begin{itemize}
\item we start from the measured flux of throughgoing muons;
\item we extrapolate this flux at low energy with the shape suggested by the HESE data;
\item \corrd{we adopt the smooth spectrum of equation \eqref{stdflux}. In this manner we determine the \lq\lq universal\rq\rq \ cosmic neutrinos spectrum.} 
\item we use the universal spectrum, neutrino oscillations and experimental constraints to predict the flux $\phi_\tau$ and $\phi_e$.
%we predict the fluxes $\phi_\tau$ and $\phi_e$, using theoretical and experimental considerations. 
\end{itemize}
\corrd{The last step of this procedure concerns the following two sections.} 
In other words, we are going to test whether \textit{for some production mechanisms}
the assumption of a universal spectrum 
agrees with HESE.

%We neglect the uncertainty on the spectral index due to the fact that the normalization and the spectral are strongly anti-correlated, as can be noticed from Fig.1 of \cite{hese} and Fig.6 of \cite{muon}. It follows that the expected number of events, that is the relevant quantity for the analysis proposed in this paper in Sec.\ref{sec:likhese}, is affected much more by the uncertainty on the normalization than by the uncertainty on the spectral index. 
%In order to test the stability of our procedure we consider also another 2 power-law flux, in which the difference between the spectral indexes is reduced. Namely we consider the greater spectral index, within 1$\sigma$, suggested by throughgoing muons $\alpha=2.26$ and the smaller spectral index, within 1$\sigma$, suggested by HESE $\alpha=2.63$. This spectrum is defined by:
%\begin{equation}
%\footnotesize
%\frac{d\phi_\ell^{\mbox{\tiny min}}}{dE}=N_\ell \ \frac{0.75 \times 10^{-18}}{\rm GeV \ cm^2 \ sec \ sr} \left[ \left(\frac{E}{\rm 100 \ TeV}\right)^{-2.63}+\left(\frac{E}{\rm 100 \ TeV}\right)^{-2.26} \right]
%\label{stdflux2}
%\end{equation}
%and it is represented in Fig.\ref{fig:brokenflux} with a gray dotted line. 

%\subsection{$\phi_e$ and $\phi_\tau$ in pion decay scenario}

%\textcolor{red}{INSERIRE IL RISULTATO. SPOSTARE LA PARTE SU AREE EFFETTIVE}

%\section{Test of other production mechanisms}

\subsection{The flux of $\nu_\tau$}
\label{tau:theory}
%\begin{quote}
%Assuming that neutrino oscillations are valid the fluxes of $\nu_e$ and $\nu_\tau$ must obey to the following conditions. 
%\end{quote}
The most plausible mechanism \corr{of} high energy neutrino production is the pion decay scenario, that yields $\phi_e \simeq \phi_\tau \simeq \phi_\mu$. 

Despite the popularity of this hypothesis, in the following we choose to adopt a more conservative and unbiased position, i.e.\ we assume that the mechanism of production is unknown. Therefore, we perform a test on the flavor composition to verify what is the astrophysical scenario that is in better agreement with the observations. 

To begin with, let us discuss the general constraints that come from theoretical and experimental considerations. 

\subsubsection{Constraints from neutrino oscillations}
\label{tau:th}
The assumption of this subsection is \corrd{just} that:
\begin{center}
\lq\lq We believe in \corrand{three-flavor} neutrino oscillations\rq\rq
\end{center}
%\begin{quote}
%In the hypothesis that neutrino oscillations are valid the flux of $\nu_\tau$ must obey to the following constraint.
%\end{quote}
The only expectation we have on the production mechanism of neutrinos is that no $\nu_\tau$ are produced at the source. \corrd{This applies to any reasonable astrophysical scenario.}
%following the general picture of standard neutrino production mechanisms in astrophysical environments. 
Therefore, the flavor composition at the source, defined as $\xi_\ell^0=\phi_\ell^0/\phi^0$ (where $\phi^0$ denotes the all-flavor neutrino flux at the source), is given by:
\begin{equation}
(\xi_e^0:\xi_\mu^0:\xi_\tau^0)=(x:1-x:0) \qquad x \in [0,1]
\label{eq:x}
\end{equation}
We do not distinguish between neutrinos and antineutrinos \corr{for the moment}; we just consider the total flux for each flavor $\ell$. Using this notation we have that:
\begin{itemize}
\item $x=1$ denotes the neutron decay scenario;%, in which only electron antineutrinos are produced by the well known process $n \rightarrow p + e^- +  \bar{\nu}_e $;
\item $x=1/3$ denotes the pion decay scenario;%, in which neutrinos are produced by the following chain of decays: \end{equation}\pi^+ \rightarrow \mu^+ + \nu_\mu \rightarrow \mu^+ +e^+ + \nu_\mu + \bar{\nu}_\mu + \nu_e\end{equation} or \end{equation}\pi^- \rightarrow \mu^- + \bar{\nu}_\mu \rightarrow \mu^-+ e^- + \nu_\mu + \bar{\nu}_\mu + \bar{\nu}_e\end{equation}
\item $x=0$ denotes the damped muon scenario, \corr{in which muons, produced by pion decay, interact before decaying. Therefore only $\nu_\mu$ (or $\bar{\nu}_\mu$ or both) are produced.}%in which only the first decay of the chain is allowed, as muons interact before decaying. 
\end{itemize}

Since in section \ref{muspec} we have defined the two power-law spectrum of muon neutrinos, it is interesting to compute the ratio between the flux $\phi_\tau$ and the flux $\phi_\mu$ after neutrino oscillations.
The ratio is given by the following expression:
\begin{equation}
R_{\tau \mu}=\frac{P_{e\tau} x + P_{\mu\tau} (1-x) }{P_{e\mu} x + P_{\mu\mu} (1-x)} 
\end{equation}
%considering $\xi_e \in [0,1]$ and $\xi_\mu = 1-\xi_e$.
that using the natural parametrization becomes:
\begin{equation}
R_{\tau \mu}=\frac{2+3P_0 (1-3x)-6P_1 x-6P_2(1-x)}{2+3P_0(1-3x)-6P_1(1-2x)+6P_2(1-x)}
\label{eq:rtaumu}
\end{equation}
Since $P_1,P_2$ \corrd{are small}, this ratio is equal to:
\begin{equation}\corrd{R_{\tau \mu} \simeq 1 + \mathcal{O}(P_1) + \mathcal{O}(P_2)}\end{equation} 
for every mechanism of production.

\begin{figure}[t]
\includegraphics[scale=0.45]{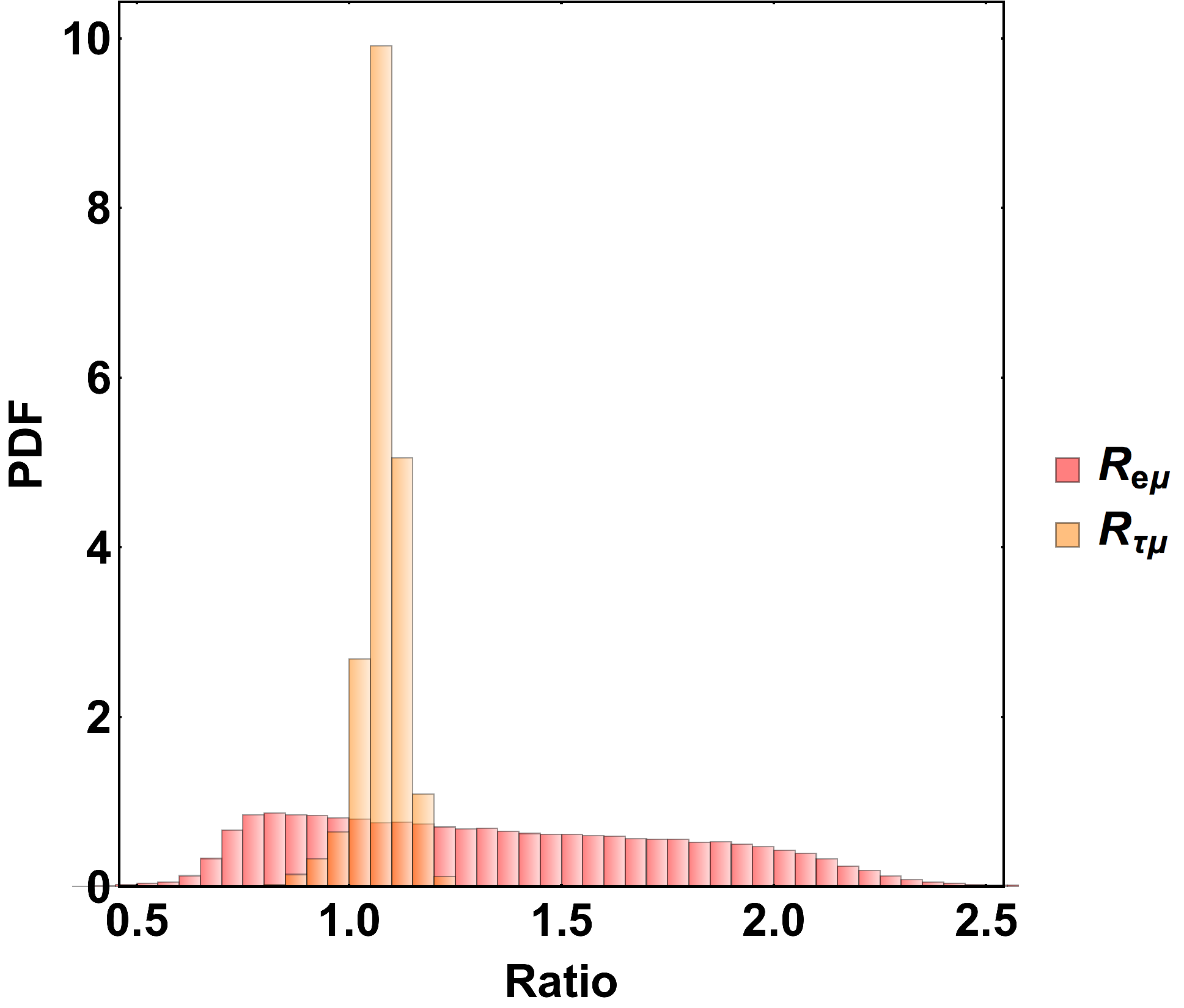}
\caption{\textit{The distributions of $R_{\tau \mu}$ (yellow) and of $R_{e \mu}$ (red) obtained from all neutrino production mechanisms (uniformly weighted) which neglect tau (anti-)neutrinos at the source.}}
\label{fig:ratios}
\end{figure}

Randomly sampling $x$ in $[0,1]$ according to a uniform distribution, so as to consider also mixed mechanisms of production, we obtain the distribution of $R_{\tau\mu}$ represented in figure \ref{fig:ratios} by orange bars. The ratio between the flux of $\nu_\tau$ and the flux of $\nu_\mu$ is, in good approximation, a Gaussian function. The best fit value and the 68\% CL interval are:
\begin{equation}
R_{\tau\mu} = 1.08 \pm 0.05
\label{eq:rtaumures}
\end{equation}
\corrd{Therefore, the amount of cosmic $\nu_\tau$ is the same of $\nu_\mu$, to a very good approximation.}
This result takes into account \corrd{also} the uncertainties on neutrino oscillations. 

\corrand{Combining equation \eqref{eq:rtaumures} with the normalization of $\phi_\mu$ (see equation \eqref{eq:likmu}), we find:
\begin{equation}
N_\tau^{\mbox{\tiny th}}=1.62 \pm 0.51 
\label{ntau:th}
\end{equation}
\corrand{This} means that the theory it is sufficient to firmly constrain the flux of tau neutrinos and antineutrinos.

In the next subsection we analyze whether it is possible to improve the knowledge of $\phi_\tau$, using informations provided by the observations.}

%is about the same of the flux of $\nu_\mu$, within a 10\%  (see Fig.\ref{fig:ratios}):

\subsubsection{Constraints from observations: double pulses}
As we have seen in the previous subsection, tau neutrino production at the source is neglected in any plausible neutrino production scenario, but, thanks to neutrino oscillations, we expect the $\nu_\tau(\overline \nu_\tau)$ flux to be approximately equal to the flux of $\nu_\mu(\overline\nu_\mu)$, regardless of the mechanism of production of high energy neutrinos \corrd{(see equation \eqref{eq:rtaumures})}.

Unfortunately, it is quite hard to measure the flux of $\nu_\tau$ directly because, until some hundreds of TeV, tau neutrinos do not produce a peculiar signature in neutrino telescopes. With increasing energy the possibility to tag a $\nu_\tau$ increases, since the first vertex of interaction, in which the $\tau$ is created, and the second vertex of interaction, in which the $\tau$ decays, \corrd{become distinguishable}. This process  has been proposed by the IceCube collaboration in \cite{icetau} and it is called \lq\lq double pulse\rq\rq.

In \cite{2pulse} Palladino et al.~derived accurate
parametrizations of various effective areas relevant for the analysis. The effective area of double pulse is given by:
\begin{equation}
A^{\text{2P}}_\tau= \bar{A}_{\text{2P}}
  \left(\frac{E_\nu}{\text{1\ PeV}} \right)^{\!\beta} \ \exp\left(-\frac{E_\text{min}}{E_{\nu}} \right) 
 \label{aeff:2bang}
\end{equation}
 with
\[\left\{\begin{array}{l}\bar{A}_\text{2P}=\SI{2.33}{\square\meter}\\[1ex] \beta=0.455\\[1ex]  E_\text{min}=\SI{0.5}{PeV}\end{array} \right.\]
\corr{This analytical parametrization reproduces well the effective area of double pulses provided by the IceCube collaboration in \cite{icetau}.}

Using our benchmark flux reported in equation \eqref{stdflux}, the expected number of double pulse events can be estimated as:
\begin{equation}
\corrand{\mathcal{R}_{\mbox{\tiny 2P}}}= 4 \pi  \mbox{T} \int_0^\infty \frac{d\phi_\tau}{dE} \ A^{\text{2P}}_\tau \ dE 
\label{eq:nev}
\end{equation}
where T is the exposure time. Considering 5.7 years of exposure the expected number of events is:
\begin{equation}
\corrand{\mathcal{R}_{\mbox{\tiny 2P}}}(N_\tau)= 0.44 \times N_\tau 
\label{eq:rate2p}
\end{equation}

Up to now no double pulse events have been observed by the IceCube collaboration; it is then possible to associate a probability to the normalization $N_\tau$, given by the lack of observations. Using Poissonian statistics, the probability to observe zero events is given by:
\begin{equation}
\mathcal{L}_{\tau}^{obs} \propto \exp\left[-\corrand{\mathcal{R}_{\mbox{\tiny 2P}}}(N_\tau)\right]
\label{liktauobs}
\end{equation}

%In order to evaluate what is the impact of the uncertainty on the spectral index (at high energy) and the effect of an energy cutoff we have repeated the calculation of $R_{\mbox{\tiny 2p}}$ as a function of these two parameters. The result is illustrated in Fig.\ref{}. We take into account that the normalization and the spectral index are anti-correlated, as illustrated in Fig.6 of \cite{muon} and we have considered an exponential energy cutoff. 

%\begin{figure}[t]
%\includegraphics[scale=0.6]{n2p.png}
%\caption{Expected number of double pulse events in 5.7 years, as a function of the spectral index of the high energy spectrum }
%\end{figure}

\subsubsection{Theory and observations}
Combining theoretical expectations, due to neutrino oscillations, with the most recent measurements of the flux of $\nu_\mu$ and with the absence of double pulse events, it is possible to put a strong constraint on the expected flux of $\nu_\tau$ with cosmic origin. 

The likelihood of $\phi_\tau$, apart from a normalization factor, is \corrd{obtained using \corca{equations} \eqref{eq:rtaumu} and \eqref{liktauobs} as follows:}
\begin{equation}
\mathcal{L}_\tau(N_\tau) \propto \corrand{N_\tau} \int_0^\infty R_{\tau\mu}\left(\frac{N_\tau}{N_\mu} \right) \ \mathcal{L}_\tau^{obs}(N_\tau) \ \mathcal{L}_\mu (N_\mu) \ \corrand{\frac{dN_\mu}{N_\mu^2}}
\end{equation}
where $\mathcal L_\mu(N_\mu)$ is a Gaussian function with mean value equal to 1.5 and standard deviation equal to 0.5 (see equation \eqref{eq:likmu}). \corrand{Note the Jacobian 
$R_{\tau\mu}(y) dy =  R_{\tau\mu}(N_\tau/N_\mu) N_\tau \, d N_\mu/N_\mu^2$ in the previous integral.}
The \corrd{resulting} function $\mathcal{L}_\tau(N_\tau)$ is, in good approximation, a Gaussian function, with:
\begin{equation}
N_\tau=1.48 \pm 0.54
\label{eq:ntauglobal}
\end{equation}
\corrand{This result is very similar to the one of equation \eqref{ntau:th}. This means that:
\begin{quote}
\textit{neutrino oscillations  \textbf{alone} are sufficient to constrain the flux of tau neutrinos, given the flux of muon neutrinos.}
\end{quote}
It is important to remark that the above results \textit{do not depend upon the mechanism of production}, since we take into account a generic mechanism in the computation of the function $R_{\tau\mu}$.}

\subsection{The flux of $\nu_e$ and $\bar{\nu}_e$}
\label{sec:nueantinue}
%\textcolor{red}{inserire considerazioni su pione scenario}
As already done for tau neutrinos, we can consider theoretical and experimental constraints for the flux of $\nu_e$ and $\bar{\nu}_e$ \corrd{separately}. Let us remark that the flux of $\bar{\nu}_e$ is constrained by the non observation of resonant events, which we discuss in section \ref{secglashow}.

%\huge
%\textcolor{red}{RIPRENDERE DA QUI}
%\normalsize

\subsubsection{Constraints from neutrino oscillations}
We follow the same procedure adopted in section \ref{tau:th} also for electron neutrinos and antineutrinos. In this case the ratio between $\phi_e$ and $\phi_\mu$ is given by:
\begin{equation}
R_{e \mu}=\frac{P_{ee} x + P_{e\mu} (1-x) }{P_{e\mu} x + P_{\mu\mu} (1-x)} 
\end{equation}
Using the natural parametrization it becomes equal to:
\begin{equation}
R_{e \mu}=\frac{2-6P_0 (1-3x) + 6 P_1(1-x)}{2+3P_0(1-3x)-6P_1(1-2x)+6P_2(1-x)}
\label{eq:remu}
\end{equation}
%\corr{This ratio is constrained between a minimum value $\sim \frac{2-6P_0}{2+3P}$, obtained for x=0, and a maximum value, obtained for x=1}
Also in this case we consider a \corrd{generic} mechanism of production, performing a uniform extraction of $x$ between 0 and 1. The resulting distribution of $R_{e \mu}$ is non Gaussian, as it can be noticed from figure \ref{fig:ratios} (red bars). The \corrd{mode} and the 68\% CL interval are given by:
\begin{equation}
R_{e \mu} = 0.81^{+0.85}_{-0.10} 
\end{equation}
\corrand{Combining the last result with equation \eqref{eq:likmu} we find
\begin{equation}
N_e^{\mbox{\tiny th}}=1.46^{+1.18}_{-0.62}
\end{equation}
The uncertainty on $N_e^{\mbox{\tiny th}}$ is quite large; therefore neutrino oscillations alone are not sufficient to constrain accurately $\phi_e$. This is due to the fact that,}
unlike the ratio $R_{\tau \mu}$, the ratio $R_{e\mu}$ strongly depends upon the mechanism of production. %Therefore neutrino oscillations are not sufficient to constrain $\phi_e$. 

In order to constrain $\phi_e$ \corrand{we can rely on the existing data:}
\begin{enumerate} 
\item the showers observed in HESE dataset;
\item the lack of resonant events.
\end{enumerate}
\corrand{Let us emphasize that only at this point, i.e.\ when we consider these two experimental ingredients, we can obtain indications on the mechanism of cosmic neutrino production.}
%The flux of $\nu_e$ is between 0.5 and 2.5 times the flux of $\nu_\mu$ (see Fig.\ref{fig:ratios}):

\subsubsection{Flux of $\bar{\nu}_e$: Glashow resonance}
\label{secglashow}
\corrd{The process:
\begin{equation}
\bar{\nu}_e + e^- \rightarrow W^-
\end{equation}
is called \lq\lq Glashow resonance\rq\rq \ \cite{glashow} and happens for electron antineutrinos with an energy of 
as 6.32 PeV (resonance).}
Assuming that the flux of neutrinos has no energy cutoff below 6.32 PeV, the resonant events, produced in the interaction of $\bar{\nu}_e$ with the electrons in the ice, must be observed. In several papers \cite{natpar,2pulse,glashow1,glashow2,glashow3} the possibility to discriminate the production mechanisms of high energy neutrinos using the resonant events has been investigated, since different production mechanisms produce a different amount of $\bar{\nu}_e$. 

The Glashow resonance cross section is given by:
\[
\sigma_{\text G}^\text{hadr} (E)=\cfrac{G_{\text F}^2 \ (\hbar c)^2 \ M_{\text W}^2}{3\pi}\cfrac{E\times \overline{\text{BR}}}{E_{\text G}\left[\left( \cfrac{E}{E_{\text G}}-1\right)^2+\left(\cfrac{\Gamma_{\text W}}{m_{\text G}} \right)^2\right]} 
\]
where $G_{\text F}$ is the Fermi constant, $M_{\text W} \simeq \SI{80}{GeV}$ is the mass of the $W^-$ boson, $\Gamma_{\text W}=\SI{2.085}{GeV}$ is its FWHM, and $E_{\text G}=M_{\text W}^2/2  M_e \simeq \SI{6.32}{PeV}$ is the energy at which the cross section is largest. The coefficient $\overline{\text{BR}} \simeq 20/3$ denotes the ratio between the branching ratio of the hadronic channel and the branching ratio of $W^- \rightarrow \bar{\nu}_\mu + \mu^-$. Here  \corr{we consider} the hadronic channels only, that produce a distinguishable signal in the detector (for a discussion of the leptonic ones see \cite{2pulse}). 

The expected number of events can be computed using the following general formula:
\begin{equation}
\corrand{\mathcal{R}_\ell}= 4\pi \text{T} \int_0^\infty \frac{d\phi_\ell}{dE} \ A_\ell \ dE
\label{eq:nev}
\end{equation}
where $A_\ell$ is the effective area for each flavor, T is the exposure time (fixed to 5.7 years) and the flux is given by equation \eqref{stdflux}.
%of Eq.\ref{eq:nev} to compute the expected number of events, 
For the specific case of resonant events we use the flux of $\bar{\nu}_e$. A useful approximation of the hadronic Glashow resonance effective area is obtained using the Dirac $\delta$ function, as follows:
\begin{equation}
A_{\bar{\nu}_e}^{\text{G}}(E)= 1.15 \times 10^6 \times \delta \left(\frac{E}{\SI{1}{TeV}}-6320 \right) \text{m}^2
\end{equation}

\corrd{Using the benchmark flux defined in equation \eqref{stdflux},} the expected number of resonant events, after 5.7 years of exposure, is equal to:
\begin{equation}
\corrand{\mathcal{R}_{\mbox{\tiny G}}}(N_e, \epsilon)= 2.3 \ N_e \times \epsilon 
\label{rateglashow}
\end{equation}
where the parameter that quantify the asymmetry between 
electron neutrinos and antineutrinos is simply:
\corrd{\begin{equation} \label{babilon}
\epsilon = \frac{\phi_{\bar{\nu}_e}}{\phi_{\bar{\nu}_e}+\phi_{\nu_e}}, \ \ \ \ \ 0\leq \epsilon \leq 1
\end{equation}}
The quantity $\epsilon$ is related to the mechanism of production and provides complementary information with respect to the parameter $x$ (see equation \eqref{eq:x}). Let us summarize:
\begin{itemize}
\item $\epsilon=1$ derives from neutron decay scenario, because only $\bar{\nu}_e$ are produced in this mechanism; 
\item $\epsilon \simeq 1/2$ comes from the proton-proton interaction, in which an about equal amount of $\nu_e$ and $\bar{\nu}_e$ is produced;
\item $\epsilon \simeq 1/4$ comes from the ideal $p\gamma$ mechanism ($\delta$ approximation, i.e.\ only the $\Delta^+$ resonance is produced). In more realistic scenarios, analyzed in \cite{glashow3,winter}, $\epsilon$ \corrd{is} larger that 1/4, due to the production of $\pi^-$;
\item $\epsilon=0$ is obtained in extreme scenarios, in which there are no antineutrinos at the source \corr{at all}. This happens when only $\pi^+$ are produced and only the first decay ($\pi^+ \rightarrow \mu^+  + \nu_\mu$) is allowed. For example, it could happen in an ideal $p\gamma$ mechanism, in which muons interact before decaying (damped muons scenario).
\end{itemize}

\corrand{Since no resonant events have been detected by IceCube up to now \cite{iceproceeding}}, it is possible
to associate a prior distribution to the normalization of the $\bar{\nu}_e$ flux, i.e.\ to $N_e \times \epsilon$, \corrd{related to the non observation of resonant events}. \corrd{Using the Poissonian statistics} the likelihood is given by:
\begin{equation}
\mathcal{L}_{\bar{\nu}_e}(N_e \times \epsilon) \propto \exp\left[-\corrand{\mathcal{R}_{\mbox{\tiny G}}}(N_e, \epsilon)\right]
\label{likantinue}
\end{equation}
with the condition $\epsilon \in [0,1]$.

\corrd{Finally, just as for double pulse events, we remark that 
the assumption on the low energy part of the spectrum does not affect the result, since only very high energy neutrinos contribute to the resonant events; the broken power law or the double power laws are equivalent for the purpose of estimating the 
number of Glashow resonance events.}

%It is important to remark that the expected number of resonant events depends only on the high energy part of the spectrum, i.e.\ the $\sim E^{-2.1}$ spectrum in our two power-law hypothesis. 
%, in order to consider realistic mechanisms of production. 

%B.R.(W^- \rightarrow \bar{\nu}_\mu \ \mu^-)$ is the branching ratio of the leptonic channel, that is equal to $B.R.(W^- \rightarrow \bar{\nu}_\mu \ \mu^-) \simeq 0.105$ \cite{pdg}.
%\end{equation}
%\frac{1}{B.R.(W^- \rightarrow \bar{\nu}_\mu \ \mu^-)}
%\end{equation}
%There is a prior distribution for the normalization of $\phi_{\bar{\nu}_e}$ due to the non observations of resonant events (not negligible effect for $\phi_{\bar{\nu}_e}$).

\subsubsection{The flux of $\nu_e$ + $\bar{\nu}_e$: HESE and theory}
The strongest constraint on the normalization of the $\phi_e$ flux comes from the number of showers observed with contained events (HESE). In fact, the $\phi_\mu$ flux gives a negligible contribution to the showers, whereas the flux of $\phi_\tau$ is fixed (within the uncertainty) by the theoretical and experimental constraints analyzed in section \ref{tau:theory}. This means that the degrees of freedom needed to reproduce the observed number of showers are $N_e$ and $\epsilon$.\footnote{\corr{Let us clarify that we are assuming that all $\nu_\tau$ are detected as showers. This assumption is not completely true but, even if about 20\% of tau neutrinos would produce tracks, it would affect our result at level of 
$
0.2  k_\tau/49 \simeq 3.8\%
$,
where 49 denotes the average number of showers with a plausible astrophysical origin, after subtracting the atmospherical background given in 
equation \eqref{eq:back}.}}

%\paragraph{Atmospheric background} Let us recall that the expected number of showers, due to the atmospheric background, are equal to:
%\end{equation}
%b_s= 8.8^{+4.0}_{-3.0}
%\end{equation}
%as described in Sec.\ref{sec:background}.%We use the estimation reported in Tab.4 of \cite{ice3yr} for the conventional background and the most recent estimation on the contribution of prompt neutrinos reported in Tab.1 of \cite{nugal2}. 
%We combine the informations on the background contained in Tab.4 of \cite{ice3yr}, in Tab.1 of \cite{nugal2} and in the recent proceeding of the IceCube collaboration \cite{iceproceeding}.

%The contribution of conventional atmospheric neutrinos and prompt neutrinos to HESE showers is equal to:
%\end{equation}
%R_b=7.2^{+4.1}_{-3.6}
%\end{equation}
%We denote with $\mathcal{L}_s(b_s)$ the distribution of the atmospherical background that reproduce the best fit and the 68\% CL.
%\vskip3mm \noindent
%\paragraph{Flux of $\nu_e$ + $\bar{\nu}_e$: HESE showers\\[1.5mm]}
We use the effective areas of HESE, reported in \cite{icescience} and on the IceCube website, to evaluate the expected number of events for each neutrino flavor. We compute these expectations using equation \eqref{eq:nev}.

%%%%%%%
\corrd{Using the benchmark flux given in equation \eqref{stdflux},} the expected number\corrd{s} of showers for each neutrino flavor are given by:
\begin{eqnarray}
\corrand{\mathcal{R}_e} & = & N_e [k_{\nu_e} (1-\epsilon) + k_{\bar{\nu}_e} \epsilon ] \nonumber \\
\corrand{\mathcal{R}_\mu} & = & N_\mu k_\mu \nonumber \\
\corrand{\mathcal{R}_\tau} & = & N_\tau k_\tau  \nonumber
\end{eqnarray}
where the coefficients $k_\ell$ are equal to:
\begin{equation}
k_{\nu_e}=14.7; \ \  
k_{\bar{\nu}_e}= 17.8;   \  \ 
k_\mu=1.3;   \  \ 
k_\tau= 9.3  
\end{equation}
For $\corrand{\mathcal{R}_e}$ we need to distinguish between the contribution of $\nu_e$ and $\bar{\nu}_e$, since only $\bar{\nu}_e$ can produce resonant events. Let us notice that 
\begin{equation}
k_{\bar{\nu}_e} - k_{\nu_e} > 2.3,  \  \text{see equation \eqref{rateglashow}}
\end{equation}
because in the effective areas also the leptonic channels are included, which give showers below 6.32 PeV, which are not distinguishable from those produced by deep inelastic scattering \cite{2pulse}. For $\corrand{\mathcal{R}_\mu}$ we take into account that only 20\% of events produced by muon neutrino plus antineutrinos are shower-like events, as discussed in \cite{prl}.

\corrd{Using the previous coefficients $k_\ell$, we define the likelihood $\mathcal{L}_{\mbox{\tiny HESE}}$ as follows, taking into account that the observed number of showers is $\corrand{\mathcal{R}_s}=58$:
\begin{equation}
\begin{split}
\mathcal{L}_{\mbox{\tiny HESE}} (N_e, \epsilon) \propto [b_s+N_\mu(k_\mu+ k_\tau)+\corrand{\mathcal{R}_e}(N_e,\epsilon)]^{\corrand{\mathcal{R}_s}} \times \\ 
\exp[-(b_s+N_\mu(k_\mu+k_\tau)+\corrand{\mathcal{R}_e}(N_e,\epsilon)]
\end{split}
\label{eq:likhese}
\end{equation} }

\corrd{Adding the prior distribution $R_{e\mu}$ \eqref{eq:remu}, $\mathcal{L}_{\bar{\nu}_e}$ \eqref{likantinue}, $\mathcal{L}_\mu$ \eqref{eq:likmu}, $\mathcal{L}_s$ \eqref{eq:back}, we compute the complete likelihood function of $N_e$ and $\epsilon$ as follows:
\begin{equation}
\begin{split}
\mathcal{L}_e (N_e, \epsilon) = \corrand{N_e} \int_0^\infty \corrand{\frac{dN_\mu}{N_\mu^2}} \int_0^\infty db_s\  \mathcal{L}_{\mbox{\tiny HESE}}(N_e,\epsilon) \times \\
\times R_{e\mu}\left(\frac{N_e}{N_\mu} \right) \ \mathcal{L}_{\bar{\nu}_e}(N_e,\epsilon) \ \mathcal{L}_\mu(N_\mu) \ \mathcal{L}_s(b_s)
\end{split}
\end{equation}
%where $R_s=58$ is the number of HESE showers detected in 5.7 years and
\corrd{In the previous expression} we are using $N_\mu \simeq N_\tau$ (see equation \eqref{eq:ntauglobal}), in order to simplify the calculation.}%Anyway the approximation $N_\mu \simeq 1.1 \ N_\tau$ perfectly works within the uncertainty of 35\%, that is contained into the $\mathcal{L}_\mu(N_\mu)$ function.
%The prior distributions $\mathcal{L}$ has been defined in the previous sections.  }

\begin{figure}[t]
\includegraphics[scale=0.45]{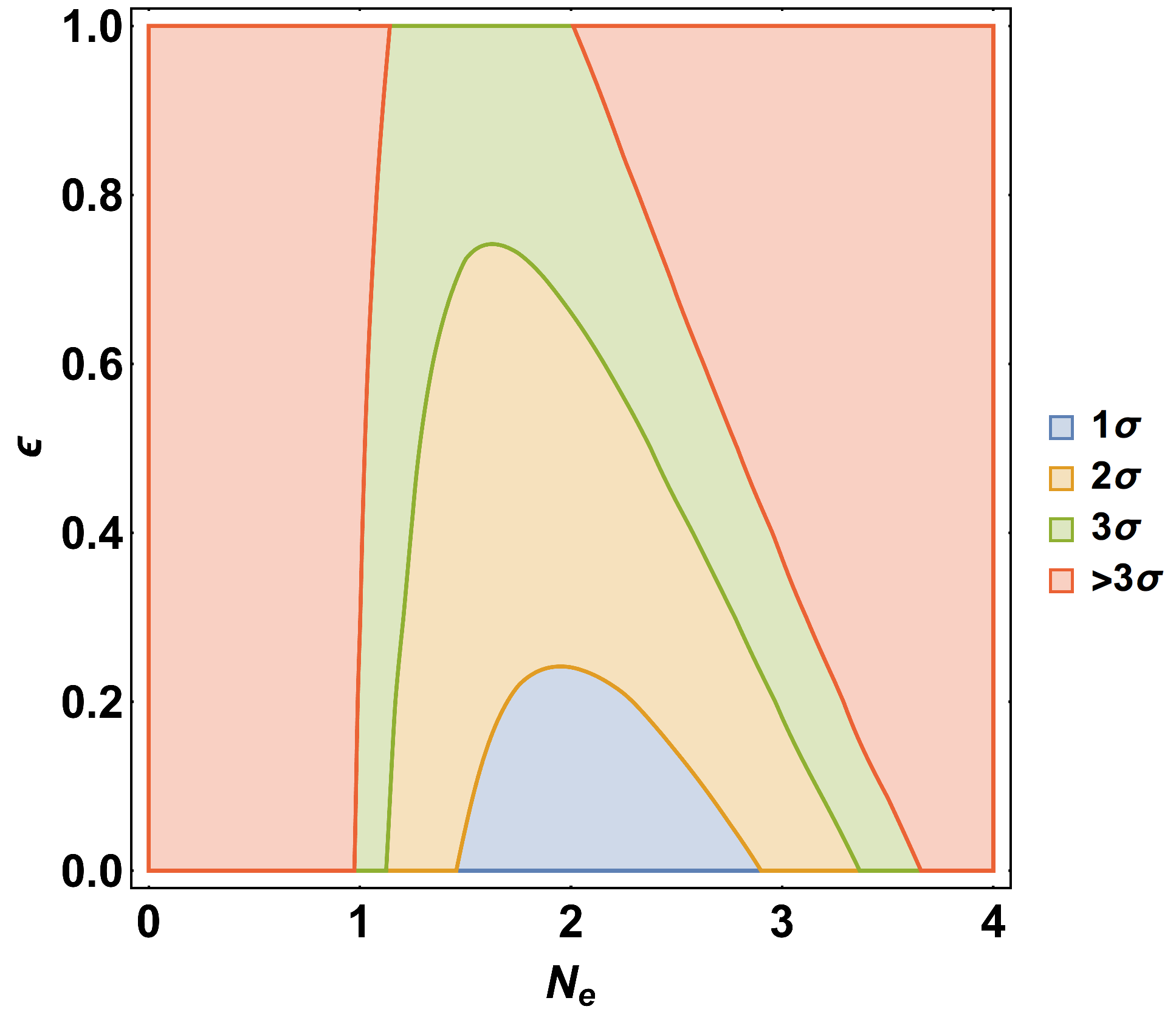}
\caption{\textit{The likelihood of $N_e$ as a function of the normalization of $\phi_e$ and of $\epsilon$, the fraction of electron antineutrinos.}}
\label{fig:2lik}
\end{figure}

The results are illustrated in figure \ref{fig:2lik}. The regions are defined using the Gaussian 2-dimensional approximation:
\begin{equation}
(1-\mbox{CL}_1) \times \mathcal{L}_e^{max} \leq \mathcal{L}_e \leq (1-\mbox{CL}_2) \times \mathcal{L}_e^{max}
\end{equation}

Marginalizing the 2-dimensional likelihood we obtain, separately, an estimate for $N_e$ and $\epsilon$:
\begin{equation}
\begin{array}{l}
N_e = 1.83 \pm 0.44   \\[1ex]
\epsilon < 0.52  \mbox{ at 90\% CL}
\label{eq:neeps}
\end{array}
\end{equation}
\corrd{We have checked that the choice between the spectrum given in equation \eqref{stdflux} and the spectrum given in equation \eqref{brokenflux} affects the previous analysis at level of few \%. The same consideration applies considering a different normalization point (within a factor 2) in the flux defined in equation \eqref{stdflux}. This demonstrates the robustness of the analyses proposed in this paper.}

%-----

%The preference for small values for $\epsilon$ comes from the non observation of resonant events. This result is valid if the neutrino spectrum has not an energy cutoff below the Glashow resonance energy. In the opposite case there is no possibility to separate the contribution of $\nu_e$ and $\bar{\nu}_e$ and, as a consequence, to extract useful information \corrd{on the} parameter $\epsilon$. \corr{On the contrary, the constraint on $N_e$ can be calculated also in this scenario and we have checked that the \corrd{impact of the cutoff} is negligible, with respect to the result obtained in this paper.}

% ----

In \corr{table \ref{tab:summary}} we summarize the results obtained in this section. 
\corrand{With these results on normalization factors of the neutrino fluxes
$N_\ell$  ($\ell=e,\mu,\tau$) and on $\epsilon$,  
we have concluded the 
definition of our model for universal spectrum of 
the cosmic neutrinos given in equation \eqref{stdflux}.}

\begin{table}[h]\caption{\textit{Summary of the normalizations of the high energy neutrino flux \corr{at Earth} defined in equation \eqref{stdflux}, divided per flavor. The parameter $\epsilon$ given in equation \eqref{babilon}  
is the fraction of electron antineutrinos with respect to the $\phi_e$ flux.\\[1mm]}}
\label{tab:summary}
\begin{tabular}{cccc}
\hline
\multirow{2}{*}{$N_e$} & \multirow{2}{*}{$N_\mu$} & \multirow{2}{*}{$N_\tau$} & $\epsilon$  \\
& & & 68\% CL - 90\% CL \\[.5mm]
\hline &&&\\[-2mm]
\corrand{$1.98 \pm 0.45 $} & $1.50 \pm 0.50 $  & \corrand{$1.48 \pm 0.54 $} & $<$\corrand{0.25} -- $<$\corrand{0.52} \\[1mm]
\hline
\end{tabular}
\end{table}

Before passing to discuss the 
predictions,  it is useful to see again figure \ref{fig:2lik} keeping in mind 
table \ref{tab:summary}. 
It can be noticed that \corrd{$N_e\simeq N_\mu \simeq N_\tau$ (expected from $\pi$ production)} is contained into the 1$\sigma$ region; moreover, a small value for $\epsilon$ is preferable.

%\subsection{Conventional background and prompt neutrinos}
%This section is necessary to introduce the likelihood for the expected number of showers. 

%\subsection{Flux of $\nu_e +\bar{\nu}_e$: HESE}
%\label{sec:likhese}
%Since the flux of muon neutrinos $\phi_\mu$ is measured and the flux of tau neutrinos $\phi_\tau$ is about $\phi_\tau \simeq \phi_\mu$, we have freedom in the fluxes of $\nu_e$ and $\bar{\nu}_e$ to reproduce the showers observed by IceCube.

%\huge
%\textcolor{red}{RIPRENDI DA QUI}%
%\normalsize

\section{Predictions and critical aspects of the model}
\corrand{Having introduced and described our model, we can assess the expectations. We will discuss in the following 
three specific instances: 
1)~we examine in section \ref{s:e1} the flavor composition of the universal spectrum defined above and compare it with
some important cases;
2)~we discuss in section \ref{s:e2} the expected number of double pulse and Glashow resonance events,
examining the uncertainties and showing their relevance;
3)~we consider in section \ref{s:e3} the angular distribution of the events and emphasize the critical importance of 
testing it for the low energy part of the spectrum, possibly, using new detectors in the Northern hemisphere.}

\subsection{Flavor composition at Earth\label{s:e1}}
\corrand{First of all, we discuss what flavor composition of the universal spectrum we obtain from our model and compare it with the theoretical expectations from some specific models for cosmic neutrino production.}
\label{sec:flavorres}
\paragraph{Theory:} Using the natural parametrization described in the first section it is trivial to compute the flavor composition expected from a theoretical standpoint for different mechanisms of production. For a generic mechanism, with initial flavor composition 
\begin{equation}
(\xi_e^0:\xi_\mu^0:\xi_\tau^0 )= (x:1-x:0)
\end{equation}
the fraction $\xi_e$ of $\nu_e$ + $\bar{\nu}_e$ after neutrino oscillations is equal to:
\begin{equation}
\xi_e(x)=x \left(\frac{1}{3}+2 P_0 \right) +(1-x) \left(\frac{1}{3}-P_0+P_1 \right)
\label{flavorteorico}
\end{equation}
where $x=1$ denotes the neutron decay scenario, $x=1/3$ the pion decay scenario and $x=0$ the damped muon scenario, as already discussed in section \ref{tau:th}. \corrd{This flavor ratio is useful because it allows a clear discrimination of the different theoretical predictions}, due to the fact that $P_{e\mu}\simeq P_{e\tau} \approx P_{ee}/2$, \corr{i.e.\ $\nu_e$ is the neutrino that mixes the least with other neutrinos.}

\paragraph{Observations:} Using the fluxes reported in \corr{table \ref{tab:summary}}, we compute the flavor composition. The normalization of the total flux \corrd{(a pure number, see equation \eqref{stdflux})} is given by:
\begin{equation}
N_{\text{tot}}= N_e+ N_\mu + N_\tau = \corrand{4.96 \pm 0.86}
\end{equation}
where the uncertainty is obtained summing in quadrature the uncertainties on the different normalizations. 
The observed flavor ratio of \corrd{$\nu_e+\bar{\nu}_e$} is thus equal to:
\begin{equation}
\corrd{\xi_e^{\mbox{\tiny obs}}} = \frac{N_e}{N_e+N_\mu+N_\tau}=\corrand{0.40 \pm 0.11}
\label{flv:res}
\end{equation}
where the uncertainty is, \corrd{as usual}, given by:
\begin{equation}
\Delta \xi_e^{\mbox{\tiny obs}}=\sqrt{\left(\frac{\Delta N_e}{N_e} \right)^2 + \left(\frac{\Delta N_{\text{tot}}}{N_{\text{tot}}} \right)^2}
\end{equation}
\corrand{The three histograms represent the predictions due to oscillations, while the gray vertical band covers the range given in equation \eqref{flv:res}.}

\begin{figure}[t]
\includegraphics[scale=0.65]{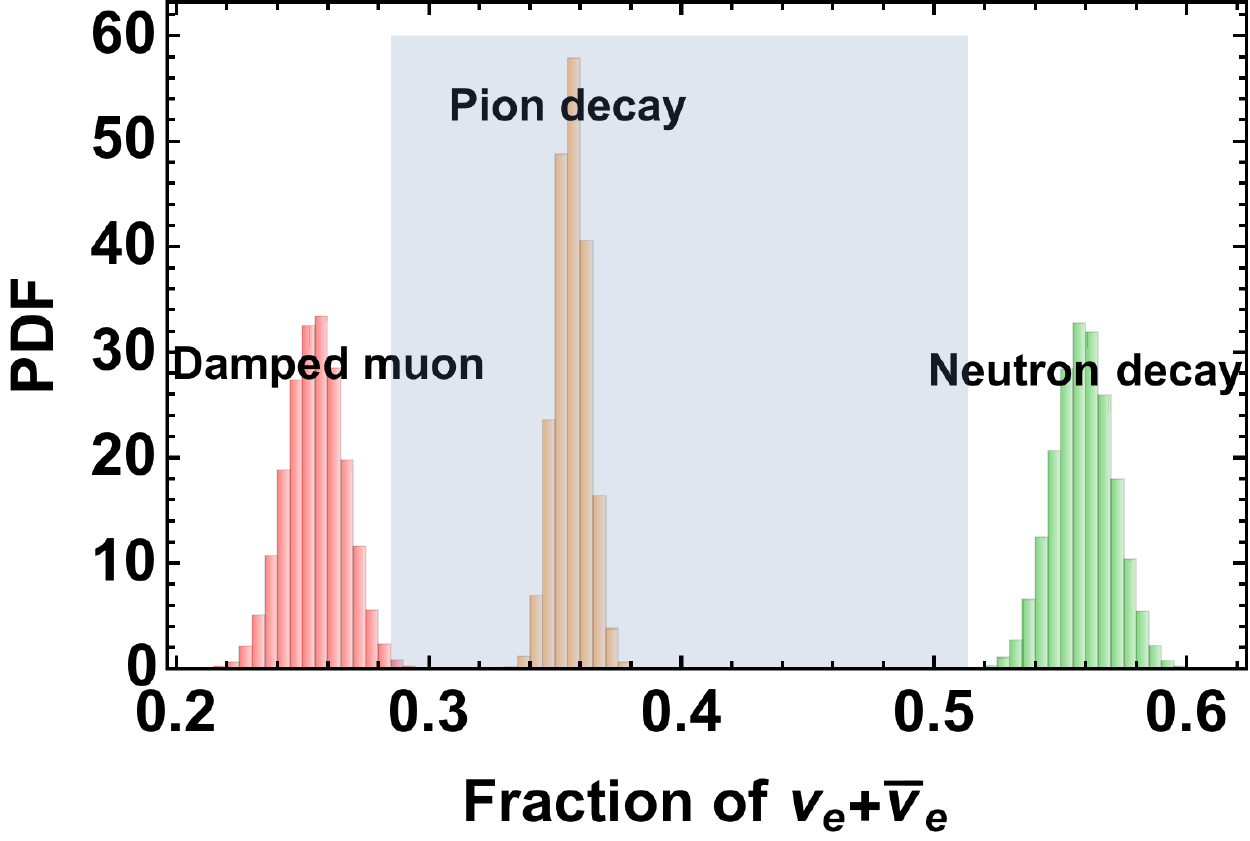}
\caption{\textit{Comparison between the theoretical flavor ratio expected from different mechanisms of production (colored histograms) and the observed one (shaded area).}}
\label{fig:flavor}
\end{figure}

\paragraph{Comparison:} The comparison between theoretical expectations \corrand{(equation \eqref{flavorteorico})} and the observed flavor ratio \corrand{(equation \eqref{flv:res})} is shown in figure \ref{fig:flavor}. 
This indicates compatibility with the pion decay scenario, that is \corr{also} the most plausible mechanism of production \corr{from a theoretical point of view}. The neutron decay scenario is excluded at about 2$\sigma$, but a stronger constraint is given by the fact that $\epsilon=1$ (i.e.\ the neutron decay scenario) is excluded at least at 3$\sigma$ (see figure \ref{fig:2lik}). On the other hand, the damped muon scenario is still compatible with the expectations within 1.5$\sigma$.

Taking simultaneously into account the flavor ratio $\xi_e$ and the preference for small $\epsilon$, we conclude that, under the hypothesis that no energy cutoff is present below $\sim 7-\SI{8}{ PeV}$, there is \corrd{an hint} for $p\gamma$ as mechanism of production. \corrd{In this scenario} high energy neutrinos are likely to be produced in the decay of $\pi^+$ and, in smaller amount, in the decay of $\pi^-$. %, that \corrd{would be} a typical situation in the $p\gamma$ interaction. 
As a consequence, the flux of  $\nu_e$ is larger than the flux of $\bar{\nu}_e$.

\subsection{Observable high energy events of new type\label{s:e2}}

\corrand{
Only the high energy part of the spectrum is relevant for the computation of double pulse events and Glashow resonance events: these events are related to the $\propto E^{-2.1}$ part of the  spectrum. There is thus no difference in expectations when we 
use the spectrum suggested by throughgoing muons, or the broken power-law spectrum of equation \eqref{brokenflux}, or the two power-law spectrum of equation \eqref{stdflux}. Let us proceed to evaluate the expectations assuming $\text T=5.7$ years of exposure.}

\paragraph{Double pulse events}
\corrand{
In section \ref{tau:theory} we have seen that $\phi_\tau \simeq \phi_\mu$, due to neutrino oscillations. We remark that it is always true for a generic production mechanism, not only for the pion decay scenario. 

This result gives 
rise to an important theoretical prediction.
Combining equations \eqref{eq:ntauglobal} and \eqref{eq:rate2p} (or similarly equations \eqref{ntau:th} and \eqref{eq:rate2p}), 
we find that the expected number of double pulse events after $\text T=5.7$ years of exposure is:
\begin{equation}
\corrand{\mathcal{R}_{\mbox{\tiny 2P}}^{\mbox{\tiny th}}=0.65\pm 0.24}
\end{equation}if we assume there is no energy cutoff.\newline
About this expectation, we find it important to remark that: 
\begin{enumerate}
\item the IceCube collaboration used a $E^{-2}$ spectrum for the calculation of double pulse events \cite{icetau}; our expectations are in 
excellent agreement with \cite{icetau} and also with \cite{2pulse}; 
\item even more importantly, 
half of the expected double pulse events are produced by neutrinos with an initial energy of 2 PeV, i.e.\ neutrinos which have been already observed by IceCube. As a consequence, tau neutrinos \textit{must be observed} in the future: it is only a matter of exposure. 
\end{enumerate}
The last consideration is very remarkable, because the observation of tau neutrinos would be the definitive proof that cosmic neutrinos have been detected.\newline
However, we have to consider that the presence of an energy cutoff could reduce the possibility to observe a double pulse event. An energy cutoff at 2 PeV, 5 PeV and 10 PeV would reduce the previous expectation to 55\%, 70\% and 85\%, respectively.}

\paragraph{Glashow resonance events}

Let us use the best fit value of $N_e$, reported in equation \eqref{eq:neeps}, with the expected number of resonant events given by equation \eqref{rateglashow} and assuming pion decay as mechanism of production (as suggested by the result of section \ref{sec:flavorres}).

%
%\begin{itemize}
%\item $R_{\mbox{\tiny G}}^{\mbox{\tiny th,pp}}=2.1 \pm 0.5 $ resonant events  in 5.7 years, assuming $pp$ interaction at the source, i.e.\ $\epsilon=0.5$;
%\item $R_{\mbox{\tiny G}}^{\mbox{\tiny th,p$\gamma$}}=1.1 \pm 0.3$ resonant events in 5.7 years, assuming ideal $p\gamma$ (only $\pi^+$ at the source) interaction, i.e.\ $\epsilon \simeq 0.25$.
%\end{itemize}

\corrand{
The number of events depends upon $\epsilon$.
Assuming $\epsilon=1/2$, namely for 
$pp$ production, this is: 
\begin{equation}
\corrand{\mathcal{R}_{\mbox{\tiny G}}^{\mbox{\tiny (pp)}}=2.28\pm 0.52}
\end{equation}
while in the case $\epsilon=1/4$, that is the idealized case of $p\gamma$ production
(or minimum value expected) this is:
\begin{equation}
\corrand{\mathcal{R}_{\mbox{\tiny G}}^{\mbox{\tiny (p$\gamma$)}}=1.14\pm 0.26}
\end{equation}
These consideration show that, if the baseline model is correct and, in particular, the
spectrum does not have a cutoff for energies much smaller than 6.32 PeV,
Glashow resonance events should be seen in the future years. \newline
Note that the preference for small values for $\epsilon$, visible from figure \ref{fig:2lik} and the relevant discussion, 
derives just from the non observation of resonant events in the current IceCube dataset.\newline
The presence of an energy cutoff much smaller than 6.32 PeV diminishes or inhibits the possibility to separate the contribution of $\nu_e$ and $\bar{\nu}_e$ and, as a consequence, to extract useful information on the parameter $\epsilon$. (On the contrary, the constraint on $N_e$ can be calculated also when a cutoff is present, and we have checked that its impact is negligible with respect to the result obtained in this paper.)\newline
We mention in passing speculative scenarios for the production of the neutrinos, with major deviations from the previous standard cases:
considering the value $\epsilon=1$ for 
the neutron decay and $\epsilon=0$ 
for the damped muon scenario with only $\pi^+$ at the source, the expected number of resonant events would become 
4.2 and 0, respectively.}

\subsection{The angular distribution of the flux\label{s:e3}}
The diffuse flux of high energy neutrinos detected by IceCube is compatible with the isotropy. \corrand{On the contrary, if we assume isotropy also for the low energy part, there is tension between the HESE analysis \cite{iceproceeding} and the throughgoing muon analysis \cite{muon}, as 
remarked in \cite{nugal2}  (even if, strictly speaking, a direct comparison is not possible, since the energy threshold of HESE is 30 TeV \cite{iceproceeding} whereas the throughgoing muons analysis, at low energy, concerns neutrinos with energy of few TeV or less \cite{muon}).
 Let us recall the argument.

The analysis of the throughgoing muons at TeV energy has been performed to identify prompt neutrinos, that are expected to follow an $E^{-2.7}$ spectrum and to be isotropically distributed.  
An astrophysical isotropic component with an $E^{-2.9}$ spectrum (as suggested by HESE) or a two power-law flux $E^{-3.5}+E^{-2.08}$ (as suggested by our model) would produce a larger excess at low energy than the one expected from prompt neutrinos \cite{nugal2}.} On the contrary, the throughgoing muon analysis did not observe any significant excess over the conventional background at low energy and an upper limit on the prompt neutrino flux has been placed. In view of neutrino oscillations, the same bound apply to tau neutrinos and similar bounds apply to electron neutrinos. 
This remarks is worth of consideration and can be regarded as an issue. It could be due to:
\begin{itemize}
\item the presence of an additional component of high energy neutrinos, observed mostly or only from the Southern hemisphere, as already suggested in \cite{nugal1,nugal2}. The multi-component model proposed in \cite{nugal2}, that predicts a Galactic contribution between 10\% and 20\%, is still compatible with the most recent experimental constraints concerning the Galactic flux, provided by ANTARES \cite{antaresgal} and IceCube
\cite{icecubegal}. However, a Galactic flux $E^{-\alpha}$, with $\alpha \in [2.4,2.7]$ is no more sufficient to explain \corr{alone} the very steep spectrum suggested by the last HESE dataset \cite{iceproceeding} that behave as $E^{-2.9}$;
\item the different backgrounds contributing to the dataset. In fact, only prompt neutrinos (discussed in subsection \ref{sec:prompt}) are relevant for the throughgoing muon analysis above 200 TeV whereas conventional neutrinos, prompt neutrinos and penetrating muons (see section \ref{sec:backhese}), are relevant for the HESE analysis, and some of them are prominent in the Southern hemisphere;
%\item \corrd{the difference to isolate the cosmic signal in HESE and throughgoing muons analysis, since the atmospheric backgrounds are not the same in these two dataset;} %a different systematic in HESE and throughgoing muon analysis \corr{not perfectly under control};
\item a larger contamination of conventional atmospheric background than expected, that could be related to an efficiency of the IceCube veto smaller than expected. \corr{This is a kind of speculative scenario that would be in agreement with the $E^{-3.5}$ component of our two power-law model, since the conventional atmospheric background (both muons and neutrinos) follows an $E^{-3.7}$ spectrum;}
\item \corrand{another change of slope between 1 TeV and 30 TeV, although it is quite hard to imagine a physical motivation that could produce this effect.}
\end{itemize}

\section{Summary and conclusions}

The findings of IceCube indicate the importance to go beyond a description of the new events based on the single power-law model, that has been used in the past to analyze the flavor composition \cite{natpar,prl,flavor4,flavor5,flavor6,flavor7}. The first analysis of the neutrino spectrum using 
two components has been performed in \cite{flavor3} but it uses only HESE. In the  
analysis proposed here,  %that the spirit of this analysis is very different from the standard analysis of the flavor composition proposed in fact, 
we combine theoretical models (mechanisms of production, neutrino oscillations) with experimental informations regarding the shape of the spectrum in different energy regions, the \corrd{absence} of double pulses, the \corrd{absence} of resonant events and the observed number of showers in HESE dataset.

Furthermore, we have \corrd{evaluated the natural parameters  
of cosmic neutrino oscillations \cite{natpar}}, %reported an updated version of the natural parametrization described in \cite{natpar}, 
using the most recent oscillation \corr{results} \cite{bari} and used these results in the analysis.

We have estimate\corr{d} the flux of each neutrino flavor, based on theoretical and experimental constraints, \corrd{assuming:}
\begin{itemize}
\item neutrino oscillations and \corrd{considering the most general} mechanism of production;
\item a two power-law spectrum, that is in agreement with throughgoing muons at high energy and with the shape suggested by HESE at low energy; 
\item the number of shower-like events observed in HESE dataset;
\item the \corrd{absence} of double pulses and resonant events \corrd{in current IceCube dataset}.
\end{itemize}
We found that such %a universal shape of the 
neutrino spectrum is in good agreement with all IceCube measurements. \corrand{Moreover we have estimated the normalizations, flavor by flavor.}

 Let us remark that \corca{the three-flavor neutrino oscillation paradigm strongly constrains the flux of tau neutrinos, that} must be very similar to the flux of muon neutrinos for every plausible mechanism of production. \corrand{From this constraint an important prediction follows; the expected number of double pulse events is about 0.7 after 6 years of exposure. Therefore, tau neutrinos \textit{must be observed} with the increase of exposure.}

We obtained a preference for the pion decay as mechanism of production of high energy neutrinos. In addition, we notice a preference for small values for the ratio $\phi_{\overline \nu_e}/\phi_e$, which could be an indication towards $p\gamma$ as a mechanism of production, if the neutrino spectra have no energy cutoff below the Glashow resonance energy. For what concerns the other mechanisms of production: \begin{itemize}
\item[-] \corrand{the neutron decay scenario is excluded at  2$\sigma$ by the flavor composition and at 3$\sigma$ by the lack of resonant events; 
\item[-] the damped muon scenario, on the contrary, is still \corrand{marginally} compatible with the data.}
\end{itemize}
\corrand{We found that, with 6 years of exposure, 2.4 resonant events are expected in the $pp$ scenario of neutrino production; in the case of the $p\gamma$ scenario the expected number of events can reach, with the same exposure, a minimum value of 1.2. Let us remark that in realistic $p\gamma$ interaction also $\pi^-$ are produced, therefore the true prediction for the rate resonant events is between 0.2 and 0.4 per year.}
%, since for a standard $pp$ mechanism we would expect $\epsilon \simeq 0.5$.

Finally, \corrand{we have remarked that it is not easy to reconcile the absence of new track events   
from the Northern sky at $\sim$ TeV with the presence of HESE showers above 30 TeV, without invoking 
a non-trivial dependence of the low energy spectrum upon the angle--i.e., some major deviation from the hypothesis of isotropy.}
With the present data it is not possible to solve this issue. The contribution of the neutrino telescopes, placed in the Northern hemisphere, is fundamental to clarify the situation. Particularly, the incoming KM3NeT has a crucial role, since it is comparable to IceCube in terms of dimension and it is complementary in terms of position.
In fact, KM3NeT will observe the Southern hemisphere using throughgoing muons and the Northern hemisphere using contained events.

\corrand{We would like to conclude stressing that the kind of analysis proposed in this paper, 
\corrd{easy and fast to implement}, is \corrd{also} very promising for the future.}

%The procedure of this paper is fast and it offers the possibility to estimate the flux of high energy neutrinos flavor by flavor. Moreover it gives the possibility to have informations of the mechanisms of production, because the value of $\epsilon$ is determined by the mechanism of production. Low values of $\epsilon$ seem to favor a $p\gamma$ mechanism of production instead of a $pp$ mechanism, in which $\epsilon \simeq 0.5$. 

%It is interesting to notice that $N_e > N_\mu \simeq N_\tau$ does not mean that there is a preference for neutrino decay as mechanism of production. In fact, in this case, the parameter $\epsilon$ would be equal to 1, but this scenario is disfavored at least at 2$\sigma$. Therefore it means that also the neutrino decay scenario is disfavored at more than 3$\sigma$. 

\end{document}